\documentclass{article}
\usepackage[preprint]{spconf}
\usepackage{amsmath,graphicx,xcolor, float}
\usepackage{enumitem}
\usepackage[explicit]{titlesec}
\usepackage[hyperfootnotes=false]{hyperref}
\usepackage{multirow}
\usepackage{soul}
\usepackage[bottom]{footmisc}
\usepackage[normalem]{ulem}
\titlespacing\section{0pt}{2pt plus 4pt minus 2pt}{0pt plus 2pt minus 2pt}
\titlespacing\subsection{0pt}{2pt plus 4pt minus 2pt}{0pt plus 2pt minus 2pt}
\titlespacing\subsubsection{0pt}{2pt plus 4pt minus 2pt}{0pt plus 2pt minus 2pt}

\title{A novel semantic compression approach for \\ultra-low bandwidth voice communication}
\name{
    \parbox{\textwidth}{
        \centering 
        Ryan Collette*, Ross Greenwood*, Serena Nicoll*
        \thanks{
            \hspace{-18pt}* Indicates equal contribution.\\
            The authors acknowledge DARPA's Strategic Technology Office for inspiring this work with their solicitation for FLexible networking Using Intelligent Dialecting (FLUID).
        }
        \vspace{.2cm} \\
        \hspace{-10pt} \textnormal{\{ryan.collette, ross.greenwood, serena.nicoll\}@str.us}
    }
}

\address{\textbf{Systems \& Technology Research}\\
600 W. Cummings Park,
Woburn, MA 01801}

\copyrightnotice{\copyright\ IEEE 2025}
                
\begin{document}
%
\maketitle

\begin{abstract}
While existing speech audio codecs designed for compression exploit limited forms of temporal redundancy and allow for multi-scale representations, they tend to represent all \textit{features} of audio in the same way.
In contrast, generative voice models designed for text-to-speech and voice transfer tasks have recently proved effective at factorizing audio signals into high-level semantic representations of fundamentally distinct features.
In this paper, we leverage such representations in a novel semantic communications approach to achieve lower bitrates without sacrificing perceptual quality or suitability for specific downstream tasks. Our technique matches or outperforms existing audio codecs on transcription, sentiment  analysis, and speaker verification when encoding at 2-$4\times$ lower bitrate---notably surpassing Encodec in perceptual quality and speaker verification while using up to 4× less bitrate. 

\end{abstract}
\begin{keywords}
neural audio codec, speech coding, \\semantic communication
\end{keywords}
\section{Introduction}
\label{sec:intro}

Although designed to maximize perceptual similarity between original and decoded waveforms, existing audio codecs often implicitly preserve low-level signal characteristics through the use of transform coding or reconstruction loss \cite{SpeechCodingReview}.
For modern speech applications that involve a fixed set of downstream tasks---such as speaker verification, emotion recognition, and transcription---these objectives may waste bits by implicitly encoding features that are irrelevant to the tasks.
In addition, low-level features have limited potential for long-term reuse.
For instance, the acoustic properties of a speaker's voice remain largely consistent across the duration of a recording, but take up additional bits in each transmitted packet when using codecs with low-level features. 
Speech processing pipelines operating over band-limited links could benefit from codecs with \textit{flexible compression objectives that adapt to downstream task requirements}, and which \textit{factor out and reuse rich persistent features} to reduce long-term temporal redundancy.

\begin{figure}[t!]
\includegraphics[width=\columnwidth]{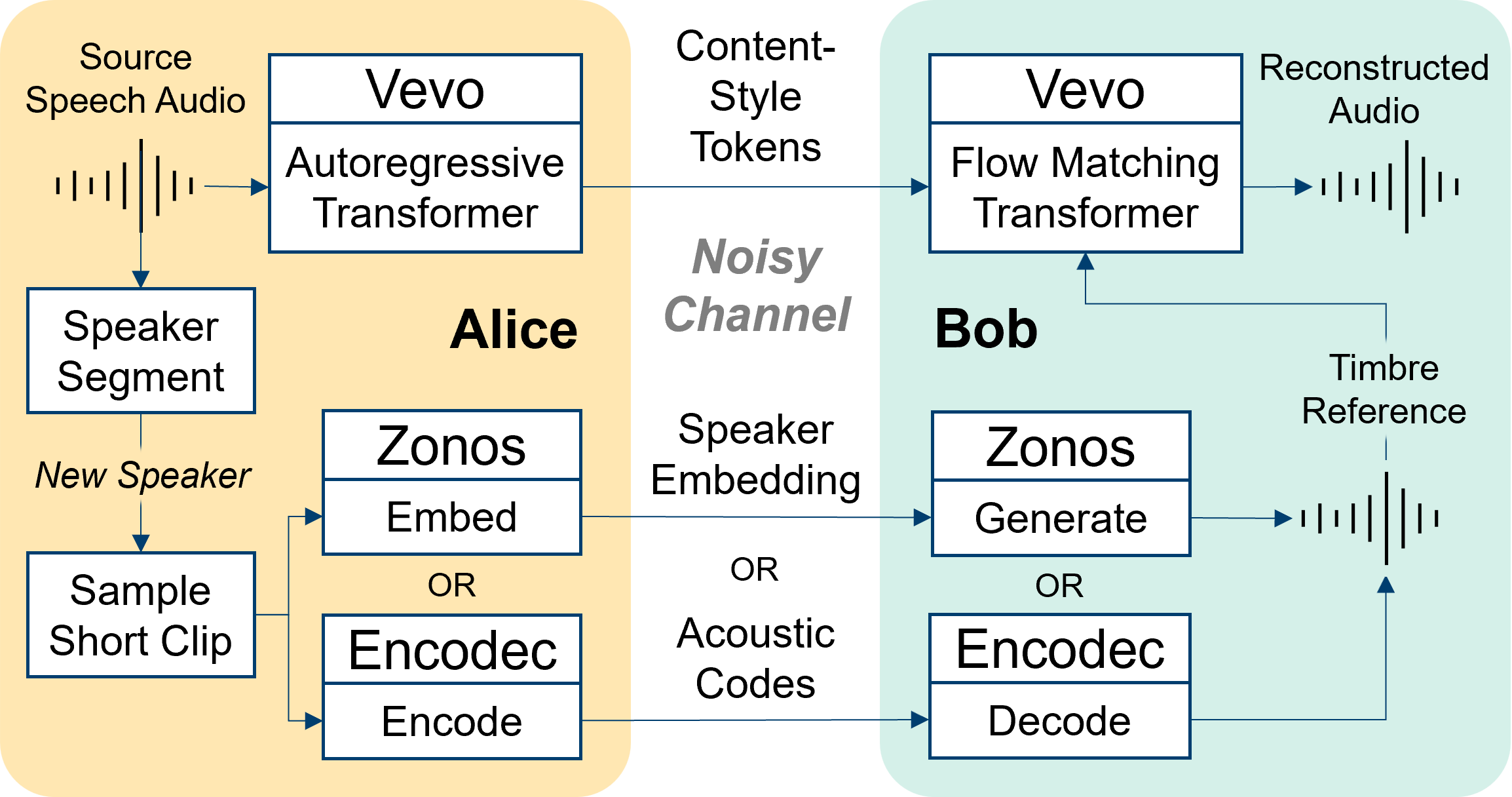}
\vspace{-16pt}
\caption{
Illustration of our approach: Alice transmits content-style tokens from Vevo continuously.
These suffice for downstream tasks of text transcription and sentiment recognition. 
When Alice detects a new speaker, she optionally transmits a compressed timbre sample (using either Zonos or Encodec) for that speaker.
Bob reconstructs a timbre Mel spectrogram from the sample (or a stock sample if none provided) and combines it with the streamed content-style tokens to regenerate the audio stream using the Vevo pipeline.
}
\vspace{-12pt}
\end{figure}

We frame these challenges as those of \textit{semantic communication}---allocating transmitted bits to represent only the semantic information that is relevant to the tasks at hand.
To address them, we propose to leverage informed strategies for three distinct levels of semantic preservation: content (what), style (how), and timbre (who). 
Factorizing the input waveform into a minimal set of semantic features for each level allows us to craft a codec that varies bitrate based on task rather than scale \cite{snac} to maintain performance at lower bitrates.

In this paper, we demonstrate the application of neural network models designed for style transfer and text-to-speech (TTS) generation to the problem of low-bitrate semantic communications. Our proposed method enables flexible communication rates while preserving semantic information relevant for downstream tasks. 
We leverage content-style tokens from Vevo \cite{vevo, amphion_v0.2, amphion} and generate encoded representations of timbre using two different methods.
Our contributions are summarized as follows:
\vspace{-6pt}
\begin{itemize}[noitemsep]
    \item We adapt generative speech models to perform semantics-preserving speech compression, minimizing bitrate and maximizing task relevant information.
    \item We introduce two schemes for timbre-preserving reconstruction using minimal compressed audio samples.
    \item We demonstrate improved performance over baselines with lower bitrates on three downstream tasks: speaker verification, sentiment analysis, and transcription. 
\end{itemize}
\vspace{-6pt}
\section{Approach}
\label{sec:approach}

Our methods focus on preserving three high-level features of speech: lexical content, prosody and linguistic content (style), and unique acoustic voice properties (timbre). Given a downstream task, we craft encoded representations of a source audio using a combination of Vevo's encoding pipeline (\S~\ref{vevoframework}) and auxiliary compression schemes (\S~\ref{encodingscheme}). We assume that both sender and receiver have prior access to Vevo and auxiliary model weights.


\subsection{Vevo Framework}
\label{vevoframework}
Vevo is a generative model capable of performing multiple types of style transfer across audio and text inputs.
Its core components are an Autoregressive Transformer and a Flow-matching Transformer, both of which accept a primary speech input as well as a reference speech sample from which to perform style conversion. 
The pipeline of interest in our approach tokenizes an input audio sample and converts it using the Autoregressive Transformer to a set of content-style tokens $Q_{s}$.
$Q_{s}$ and a Mel spectrogram $M_t$ (extracted from a speech sample) are passed to the Flow-Matching Transformer to generate a final acoustic output that retains the speech content and style of $Q_s$ and timbre reflecting $M_t$.

Our approach splits the pipeline in half and carries out the two processing steps on each side of a sender-and-receiver link.
At the sender, we compute $Q_{s}$ and a \textit{timbre sample} (described in \S~\ref{encodingscheme}) from the speech audio to be transmitted.
On the receiver, we feed $Q_s$ and an $M_t$ into Vevo's Flow-Matching Transformer to recover the original speech.

 
\subsection{Encoding Scheme}
\label{encodingscheme}

To preserve lexical and linguistic content we transmit Vevo's content-style tokens.
These tokens use a codebook of size $2^{13}$ and are generated at 50 Hz, requiring a bandwidth of just $650$ bps for realtime streaming.

To preserve acoustic voice features we transmit a \textit{timbre sample} that can be used to prompt Vevo's acoustic modeling pipeline.
By leveraging Vevo's timbre transfer capabilities a sample can be transmitted \textit{just once per speaker} and reused when decoding future utterances associated with the same speaker.
To facilitate this process, the sender performs continuous speaker diarization and transmits a new timbre sample only when a new speaker is detected.
For speakers who have already been registered, the sender transmits a short numerical identifier corresponding to the previously transmitted sample.
In our experiments, we leave speaker diarization as an exercise for the sender, and use relatively short, single-speaker audio clips to simulate individual utterances that could be part of a longer, multi-speaker recording.

Vevo's acoustic prompts are represented as 128-bin Mel spectrograms sampled at 50 Hz.
Directly transmitting these is impractical, as it would demand significantly higher bandwidth than encoding the original audio with a conventional codec.
To address this challenge, we explored two alternative methods for transmitting timbre samples: (1) switching to another audio codec for timbre transmission and (2) transmitting a speaker embedding derived from a TTS model.
We elaborate on both methods below.

\begin{figure}[ht]
    \centering
    \includegraphics[width=\columnwidth]{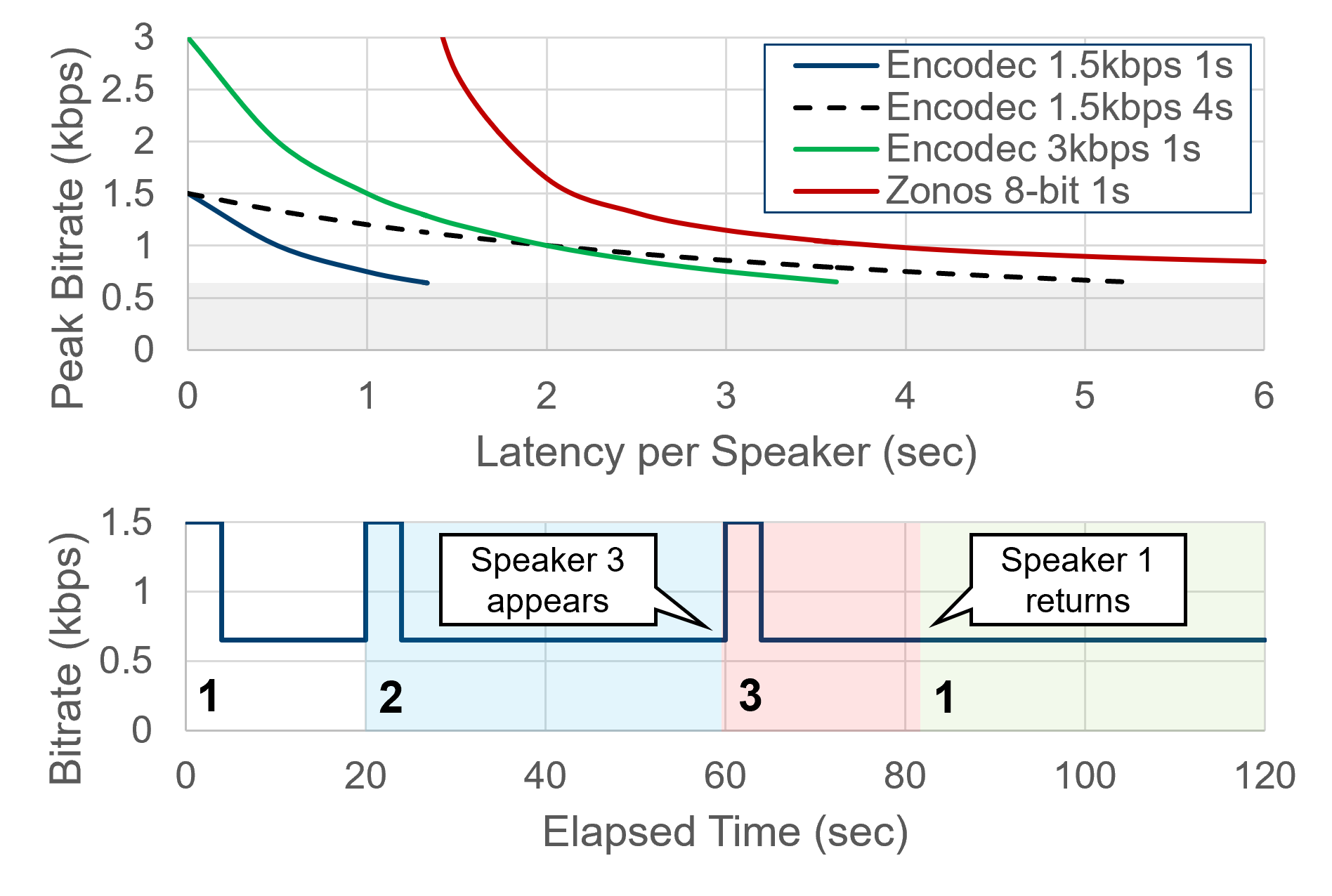}
    \vspace{-1cm}
    \caption{(Top) Trade-off between peak bitrate vs. added latency per new speaker. (Bottom) Example bitrate variability for a multi-speaker recording, using Encodec 1.5 kbps with a 4 sec timbre sample transmitted at the nominal bitrate. After all speakers have been registered, bitrate settles at 650 bps. \label{fig:bitrate_latency}}
    \vspace{-12pt}
\end{figure}

\textbf{Transmitting Timbre via Codec Switching}:
When a new speaker is detected we switch to encoding the audio signal with another codec for a duration of $d_{\mathrm{sample}}$ seconds.
During this time, the receiver can decode the complete audio signal and extract a timbre Mel spectrogram for later use.
After $d_{\mathrm{sample}}$ seconds we revert to encoding and transmitting content-style tokens only, and the receiver begins using the timbre Mel spectrogram for decoding with Vevo.
Because speakers are finite, we can operate below the codec's nominal bitrate at a finite latency cost if needed (see Fig~\ref{fig:bitrate_latency}).

\textbf{Transmitting Timbre via Speaker Embedding}:
When a new speaker is detected, a speaker embedding $z_{\mathrm{spkr}}$ is computed from an audio sample of length $d_{\mathrm{sample}}$ seconds using Zonos TTS \cite{zyphra_zonos}.
$z_{\mathrm{spkr}}$ is a 128-dimensional vector, which we quantize to 8 bits and transmit over an interval of $d_{\mathrm{transmit}}$ seconds.
On the receiver we use $z_{\mathrm{spkr}}$ and an arbitrary text prompt to generate a speech sample using Zonos, and extract $M_t$ from the resulting waveform.
During this transmission, content-style tokens are continually sent and can be decoded by the receiver if timbre information is not required.

If timbre information is required, the receiver must wait for the speaker embedding to be generated and transmitted before it can decode the audio signal.
The process introduces a latency of $L = d_{\mathrm{sample}} + d_{\mathrm{transmit}}$ seconds, where $d_{\mathrm{transmit}}$ can be adjusted to balance the trade-off between lower bitrate and higher latency, depending on the available bandwidth.
This added latency does not pose a problem for offline storage scenarios, and can still be acceptable for real-time applications that meet one or more of the following conditions: (1) the system can be ``primed" by transmitting timbre samples for all speakers in advance (2) the application can tolerate a speaker's voice loading in dynamically upon first appearance, or (3) faster-than-real-time processing can be employed to re-synchronize after periods of latency.

\section{Experiments}
\label{sec:results}

\subsection{Dataset}

We chose to use VoxCeleb1 \cite{Nagrani17}, a large-scale audio corpus of over 1M real-world utterances spanning languages, accents, and demographics.
We select an audio range from 5-20 seconds, which we regard as appropriate for simulating single utterances in typical conversations. We additionally exclude multi-speaker and non-English clips, as non-English languages are out-of-distribution for several of the downstream tasks.
Finally, we selected a random sample of 1,000 clips, balanced to include 250 examples from each of four different emotion classes used in the sentiment analysis task.

\subsection{Downstream Tasks \& Metrics}

\begin{table*}[ht!]
\centering
\begin{tabular}{l|c|c|ccc|ccccc}
& $d_{\mathrm{timbre}}$ & Bitrate & \multicolumn{3}{|c}{\textit{Downstream Task Metrics}} & \multicolumn{5}{|c}{\textit{Perceptual Quality Metrics}} \\
\textit{Model} & (sec) & (kbps) & WER$\downarrow$ & SCA & SpkRec & SpkrSim & UTMOS & NISQA & PESQ & STOI \\
\hline
Opus & N/A & 5 & 0.28 & 0.49 & 0.53 & 0.31 & 1.30 & 1.36 & 1.30 & 0.62 \\
\hline
Encodec & N/A & 1.5 & 0.20 & 0.60 & 0.63 & 0.54 & 1.26 & 2.04 & 1.45 & 0.80 \\
        & N/A & 3   & 0.12 & 0.74 & 0.74 & 0.76 & 1.30 & 2.59 & 1.90 & 0.86 \\
\hline
Vevo & & 0.65* & \multicolumn{3}{|c}{} & \multicolumn{5}{|c}{} \\
\hspace{3pt} w/o Timbre & N/A & \multirow{7}{*}{\setlength\fboxsep{0cm} \parbox{0.8cm}{\centering See\\Fig~\ref{fig:bitrate_latency}}} & 0.15 & 0.59 & 0.62 & 0.30 & 1.88 & 4.31 & 1.08 & 0.70 \\ 
\hspace{3pt} w/ Encodec@1.5kbps & 1 & & 0.15 & 0.63 & 0.68 & 0.35 & 1.61 & 4.14 & 1.12 & 0.57 \\
\hspace{3pt} w/ Encodec@1.5kbps & 4 & & 0.13 & 0.68 & 0.73 & 0.51 & 1.36 & 3.52 & 1.18 & 0.58 \\
\hspace{3pt} w/ Encodec@3kbps   & 1 & & 0.13 & 0.62 & 0.69 & 0.37 & 1.64 & 4.15 & 1.12 & 0.58 \\
\hspace{3pt} w/ Encodec@3kbps   & 4 & & 0.13 & 0.71 & 0.78 & 0.64 & 1.39 & 3.51 & 1.21 & 0.59 \\
\hspace{3pt} w/ Zonos           & 1 & & 0.14 & 0.58 & 0.72 & 0.54 & 1.72 & 4.13 & 1.14 & 0.69 \\
\hspace{3pt} w/ Zonos           & 4 & & 0.15 & 0.59 & 0.75 & 0.62 & 1.75 & 4.21 & 1.15 & 0.70 \\
\end{tabular}
\vspace{-10pt}

\caption{Comparison of metrics capturing semantics preservation and audio quality among our approach and state-of-the-art traditional and neural audio codecs. For all metrics except word error rate (WER), larger values indicate better performance. $^*$This is the \textit{default mode} bitrate, which does not include the cost of timbre transmission. \label{compareresults}}
\vspace{-12pt}
\end{table*}

\textbf{Transcription}: 
Using content-style tokens, we postulate that our method can preserve sufficient information for transcription at bitrates approaching 650 bps. We evaluate the ability of our method to preserve relevant characteristics required for good transcription by obtaining the WHISPR \cite{whispr} transcription of both source audio and reconstruction. We then compute the Word Error Rate (WER) between them: defined as $\mathrm{WER} = \frac{S + D + I}{N}$, where $S$ represents the number of substitutions, $D$ the number of deletions, and $I$ the number of insertions in the reconstructed sequence, and $N$ represents the number of words in the reference ground-truth sequence. 

\textbf{Sentiment Analysis}: Sentiment classification is done primarily based on cadence and pitch changes, which correlate with features captured in Vevo's content-style tokens. We evaluate the success of preserving speaker style using the model from \cite{Busso2008IEMOCAPIE} which leverages SpeechBrain \cite{speechbrainV1, speechbrain}. This model supports four classes: \textit{neutral}, \textit{happy}, \textit{sad}, and \textit{angry}. We use sentiment classification accuracy (SCA) as a metric, which measures the accuracy with which the reconstructed and original audios are attributed to the same emotion.

\textbf{Speaker Verification}: We evaluate the success of timbre encodings as derived in \S~\ref{encodingscheme} in preserving speaker characteristics by comparing the similarity of our input and reconstructed audio samples. For this we use a ResNet TDNN model \cite{VILLALBA2020101026} from the SpeechBrain \cite{speechbrainV1, speechbrain}. The model outputs a ``SpkRec'' score---a prediction probability between 0 and 1 for whether two audio samples come from the same speaker.

\textbf{Perceptual Metrics}:
We evaluate both subjective audio quality and speaker identity preservation in our reconstructed audio using an ensemble of metrics.
Because our audio data is noisy, we rely on no-reference metrics to measure perceptual audio quality.
Specifically, we utilize UTMOS \cite{utmos} and NISQA \cite{nisqa}, which both estimate human opinion scores without requiring clean reference audio.
Additionally, we incorporate two full-reference metrics—STOI and PESQ—to quantify low-level discrepancies between the original and reconstructed signals.
To assess the preservation of speaker identity, we calculate the cosine distance between the pyannote \cite{pyannote1, pyannote2} speaker embeddings for the original and reconstructed audio (SpkrSim).

\subsection{Performance on Downstream Tasks}
Table \ref{compareresults} demonstrates the main results of this paper, in which we compare our method to a widely used traditional audio codec (Opus) and a comparable neural audio codec approach (Encodec) at multiple bitrates. We provide results for our compression technique using the two timbre encoding strategies described in \S~\ref{encodingscheme}.

We observe that our methods consistently achieve comparable or greater performance on downstream tasks while maintaining a significantly lower bitrate. Specifically, we highlight the performance of our timbre-free approach (w/o Timbre line in Table \ref{compareresults}) at a bitrate of 0.65 kbps for sentiment classification and speaker recognition: we achieve a 59\% accuracy score for sentiment classification compared to a 60\% accuracy score using Encodec at 1.5 kbps, corresponding to a $2 \times$ decrease in bandwidth while \textit{maintaining} performance within 1\% tolerance. Similar behavior is exhibited in the case of the \textit{SpkRec} metric. Incorporating a short 4 second timbre reference using Encodec's 3 kbps codebook allows us to increase performance on all downstream tasks to a level on par with exclusively using Encodec. For our Speaker Recognition task, we \textit{increase} performance by 4\% over Encodec while maintaining a $4\times$ decrease in bitrate approaching 0.65 kbps. These results indicate that it is possible to achieve significant speedup without sacrificing performance by considering only relevant semantic information. 

Evaluating across multiple timbre encoding approaches and durations highlights that the duration of our timbre reference sample places an upper bound on the performance increase achievable for the Speaker Recognition task by switching to a more robust codebook: in the case of Encodec@1.5kbps and @3kbps, there is only a 1\% difference in performance with 1 second of timbre, while we see a 5\% increase using a 4 second reference. 

\subsection{Performance on Perceptual Metrics}


The SpkrSim metric demonstrates that reusing timbre samples can maintain speaker similarity on par with or better than Encodec while operating at a lower bitrate.
Specifically, the Vevo w/ Zonos method achieves an average speaker similarity of 0.54 with $d_{\mathrm{timbre}}=1$, matching Encodec’s performance at 1.5 kbps.
For Vevo w/ Encodec, speaker preservation exhibits greater sensitivity to the duration of the timbre sample than the bitrate at which it was encoded, with longer timbre samples required to maintain comparable performance.
Naturally, increasing the duration of the timbre sample generally enhances speaker similarity.
We remark that for the Vevo w/ Zonos approach, this adjustment increases latency but not the overall bits sent.

Across all configurations, our methods outperform Encodec at 3 kbps on UTMOS and NISQA metrics, reflecting superior audio quality in the generated outputs.
Vevo w/ Zonos outperforms Vevo w/ Encodec, likely because distortions introduced by Encodec at low bitrates are retained in the transmitted timbre sample.
Our methods score lower on full-reference metrics such as PESQ and STOI, indicating greater alterations to the signal’s original content.
This is consistent with the fact that our outputs are synthesized using a speech transfer model that was not trained to enforce preservation of low-level signal characteristics.

\subsection{Noise Resilience Analysis}

We evaluate the resilience of our encoding scheme to perform the same downstream tasks as above in the presence of increasing levels of noise. We encode audio using our best performing method in Table~\ref{compareresults} (Vevo w/ Encodec@3kbps, $d_{\mathrm{timbre}}=4$), and randomly flip bits in the transmitted tokens at rates between $10^{-3}$ and $10^{-1}$. Table \ref{fig:noiseanalysis} describes the scores for different noise levels. 

\begin{table}[h]
\centering
\begin{tabular}{c|ccc}
\hline
Noise & WER & SCA & SpkRec \\
\hline
$10^{-3}$ & 0.14 & 0.71 & 0.78\\
$10^{-2}$ & 0.32 & 0.63 & 0.77\\
$10^{-1}$ & 1.21 & 0.29 & 0.64\\ 
\hline
\end{tabular}
\caption{Degradation of downstream task metrics with increasing bit-level noise in our approach when using Vevo w/ Encodec@3kbps, $d_{\mathrm{timbre}}=4$. 
\label{fig:noiseanalysis}}
\end{table}

With $10^{-3}$, or 0.1\%, of bits flipped within the encoded signal, we observe no difference in reported metrics compared to the un-noised signal. This suggests that our approach is resilient to expected levels of noise in both traditional and constrained channels. Degradation at the 1\% bit-flip level is noticeable, but our sentiment classification and speaker similarity metrics maintain superior performance over Opus and Encodec and our WER falls within acceptable bounds. We observe significantly degraded performance on all tasks with 10\% of the signal noised.

\section{Conclusions}
\label{sec:conc}

We introduce a novel approach to compressing speech in ultra-low bandwidth regimes that leverages semantic audio representations produced by generative models.
By transmitting a subset of semantically relevant tokens, our approach achieves up to $4 \times$ lower bitrate while maintaining perceptual quality and performance on specific downstream tasks.
Furthermore, by transmitting and reusing auxiliary timbre encodings, our method can maintain speaker fidelity at similar bitrates in applications that can tolerate intermittent latency or short intervals of inaccurate voice reconstruction.
We note that for long audio streams with a limited numbers of speakers, these periods of disruption account for a small fraction of the overall audio signal.

Our approach has several limitations compared to general-purpose speech codecs. Notably, we do not address the issue of encoding overlapping speakers, which could be mitigated by separating and encoding individual speech segments on the sender side at the cost of increased bitrate during these intervals.
Additionally, errors in one-time timbre transmission result in permanent inaccuracies in voice reconstruction, highlighting the need for error recovery or periodic updates. 
Furthermore, our timbre encoding approach can cause latency or momentary disruptions during real-time streaming.
We believe such limitations can be acceptable in ultra-low bandwidth applications, and that our techniques indicate a promising direction for future research of speech compression.

\vfill\pagebreak

\bibliographystyle{IEEEbib}
\bibliography{refs}

\end{document}